\documentclass[aps,prl,preprint,superscriptaddress,onecolumn]{revtex4-1}

\usepackage{amsmath,amssymb,graphicx}
\usepackage{xcolor}
\usepackage[colorlinks=true,urlcolor=blue,linkcolor=blue,citecolor=blue,bookmarks=false]{hyperref}

\begin{document}

\title{Ferromagnetic resonance with magnetic phase selectivity by means of resonant elastic x-ray scattering on a chiral magnet}

\author{S. P\"ollath}
\affiliation{Institut f\"ur Experimentelle Physik, Universit\"at Regensburg, D-93040 Regensburg, Germany}

\author{A. Aqeel}
\affiliation{Physik-Department, Technische Universit\"at M\"unchen, D-85748 Garching, Germany}

\author{A. Bauer}
\affiliation{Physik-Department, Technische Universit\"at M\"unchen, D-85748 Garching, Germany}

\author{C. Luo}
\affiliation{Helmholtz-Zentrum Berlin f\"ur Materialien and Energie, D-12489 Berlin, Germany}
\affiliation{Physik-Department, Technische Universit\"at M\"unchen, D-85748 Garching, Germany}

\author{H. Ryll}
\affiliation{Helmholtz-Zentrum Berlin f\"ur Materialien and Energie, D-12489 Berlin, Germany}

\author{F. Radu}
\affiliation{Helmholtz-Zentrum Berlin f\"ur Materialien and Energie, D-12489 Berlin, Germany}

\author{C. Pfleiderer}
\affiliation{Physik-Department, Technische Universit\"at M\"unchen, D-85748 Garching, Germany}
\affiliation{Munich Center for Quantum Science and Technology (MCQST), Schellingstr. 4, D-80799 M\"unchen}

\author{G. Woltersdorf}
\affiliation{Institut f\"ur Physik, Universit\"at Halle-Wittenberg, D-06120 Halle (Saale)}

\author{C.H. Back}
\email[e-mail:]{christian.back@tum.de}
\affiliation{Institut f\"ur Experimentelle Physik, Universit\"at Regensburg, D-93040 Regensburg, Germany}
\affiliation{Physik-Department, Technische Universit\"at M\"unchen, D-85748 Garching, Germany}
\affiliation{Munich Center for Quantum Science and Technology (MCQST), Schellingstr. 4, D-80799 M\"unchen}


\date{\today}

\begin{abstract}
Cubic chiral magnets, such as Cu$_{2}$OSeO$_{3}$, exhibit a variety of non-collinear spin textures, including a trigonal lattice of spin whirls, so-called skyrmions. Using magnetic resonant elastic x-ray scattering (REXS) on a crystalline Bragg peak and its magnetic satellites while exciting the sample with magnetic fields at GHz frequencies, we probe the ferromagnetic resonance modes of these spin textures by means of the scattered intensity. Most notably, the three eigenmodes of the skyrmion lattice are detected with large sensitivity. As this novel technique, which we label REXS-FMR, is carried out at distinct positions in reciprocal space, it allows to distinguish contributions originating from different magnetic states, providing information on the precise character, weight and mode mixing as a prerequisite of tailored excitations for applications.
\end{abstract}

\maketitle

Ferromagnetic resonance~(FMR) measurements represent a well-established technique for the study of systems with collinear magnetization~\cite{1948_Kittel_PhysRev}, allowing to extract information on the magnetic energy landscape and material-specific parameters such as the effective magnetization, the Land\'{e} $g$ factor, or the magnetic damping constant $\alpha$. In systems with non-collinear spin textures, measurements of resonant microwave excitations exhibit very complex spectra, where the identification of specific modes proves to be prohibitively difficult. However, in view of new technological developments such as antiferromagnetic spintronics or use of quantum magnetism, the precise identification of specific modes will be of great importance~\cite{PhysRevLett.97.257202}. 

The cubic chiral magnets MnSi, Fe$_{1-x}$Co$_{x}$Si, and Cu$_{2}$OSeO$_{3}$ represent excellent showcases
for the inherent complexity of materials with technological potential. These materials host long-wavelength helimagnetic order including a trigonal lattice of topologically non-trivial spin whirls, the so-called skyrmion lattice~\cite{2009_Muhlbauer_Science, 2010_Yu_Nature, 2011_Yu_NatMater, 2012_Seki_Science, 2013_Nagaosa_NatNanotechnol}. Their microwave excitations have been studied by means of coplanar waveguides and cavities~\cite{1977_Date_JPhysSocJpn, 2010_Kobets_LowTempPhys, 2012_Koralek_PhysRevLett, 2012_Onose_PhysRevLett, 2015_Schwarze_NatMater}. In the helimagnetic states, two collective spin-precessional modes, denoted $+q$ and $-q$, are observed. In the skyrmion lattice state three eigenmodes exist, namely a clockwise and a counter-clockwise gyration mode as well as a breathing mode~\cite{2012_Mochizuki_PhysRevLett}. The remarkably detailed understanding of the cubic chiral magnet allows to assign these modes in the experimental spectra by comparing their resonance frequency and spectral weight, as well as the evolution of the latter as a function of temperature and field, to results of analytic calculations and micromagnetic simulations solving the Landau--Lifshitz--Gilbert equation taking into account dipolar interactions~\cite{2015_Schwarze_NatMater}. Such an in-depth understanding, however, may not be available when investigating novel materials~\cite{2015_Kezsmarki_NatMater, 2017_Nayak_Nature, 2017_Kurumaji_PhysRevLett} or when phenomena such as metastable states, glassy textures, phase coexistence, topological transitions, or pronounced history dependencies play a role~\cite{2010_Munzer_PhysRevB, 2012_Bauer_PhysRevB, 2016_Oike_NatPhys, 2016_Okamura_NatCommun, 2016_Karube_NatMater, 2017_Karube_PhysRevMater, 2018_Bauer_Book, 2018_Yu_Nature, 2018_Chacon_NatPhys}.

In this work we present a novel technique, called REXS-FMR, which combines the excitation of collective modes by means of a coplanar waveguide with the detection by means of the scattered intensity in magnetic resonant elastic X-ray scattering (REXS). The intensity of a crystalline Bragg peak or magnetic satellites may be studied, permitting to clearly identify the hosting magnetic state of a given excitation via the scattering pattern in reciprocal space. As a point of reference, we demonstrate the potential of REXS-FMR using the insulating cubic chiral magnet Cu$_{2}$OSeO$_{3}$, which was studied previously both by means of REXS~\cite{2014_Langner_PhysRevLett, 2016_Zhang_PhysRevB, 2016_Zhang_NanoLett, 2016_Zhang_ApplPhysLett, 2017_Zhang_NatCommun, 2018_Zhang_ProcNatlAcadSciUSA} and standard microwave spectroscopy~\cite{2012_Onose_PhysRevLett, 2013_Okamura_NatCommun, 2015_Schwarze_NatMater, 2015_Ogawa_SciRep, 2016_Seki_PhysRevB, 2017_Stasinopoulos_SciRep, 2017_Weiler_PhysRevLett}.

Our present study was carried out on the beamline PM2 at BESSY~II with the VEKMAG end station \cite{Noll:MEDSI2016-WEPE38}. The sample was a single-crystal cuboid of Cu$_{2}$OSeO$_{3}$, cut from an ingot grown by means of chemical vapor transport, with dimensions of $1.8\times0.5\times0.5~\mathrm{mm}^{3}$ and edges oriented parallel to $[110]$, $[001]$, and $[\bar{1}10]$. One of the surfaces normal to $[001]$ was mechanically polished. The latter facing top, the sample was placed into the gap of a coplanar waveguide with a gap width of 1~mm. The sample slightly protrudes the top surface of the waveguide resulting in microwave excitation that comprises both in-plane and out-of-plane components. Typical excitation fields are of the order of 3~$\mu$T to 10~$\mu$T, depending on the excitation frequency. For the REXS measurements, the energy of the circularly polarized photons is tuned to the Cu L$_{3}$ edge (931~eV). Note that the element specific character of REXS is also inherited to REXS-FMR. Further note that only in resonant X-ray scattering the crystallographically forbidden Bragg peak at $2\theta \approx 96.5^\circ$ is observed~\cite{1994_Templeton_PhysRevB, 2012_Dmitrienko_PhysRevLett, 2016_Zhang_PhysRevB, 2016_Zhang_NanoLett}.

\begin{figure}
	\includegraphics[width=0.7\linewidth]{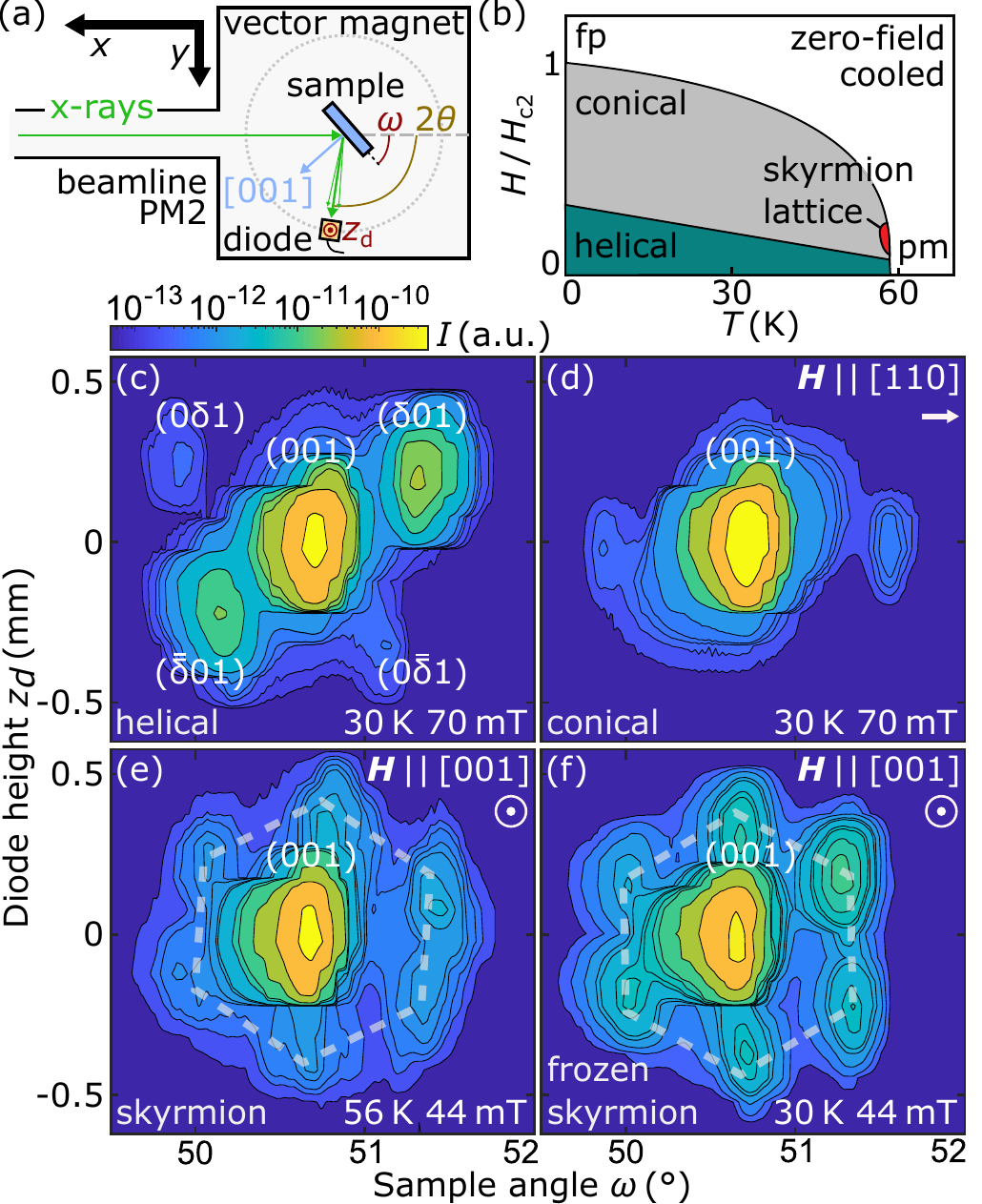}	
	\caption{\label{figure1}Setup and typical REXS data. (a)~Schematic view of the experimental setup. (b)~Schematic zero-field cooled magnetic phase diagram of Cu$_{2}$OSeO$_{3}$. (c)~Reciprocal space map around the $(001)$ Bragg peak in the helical state. (d)~Reciprocal space map in the conical state. The magnetic field is applied along the $[110]$ axis, i.e., within the sample plane. \mbox{(e),(f)}~Reciprocal space map in the skyrmion lattice state. The magnetic field is applied along the $[001]$ axis, i.e., perpendicular to the sample plane. Under field cooling, the skyrmion lattice may be metastable frozen-in to lower temperatures.}
\end{figure}

The geometry of the experiment is illustrated in Fig.~\ref{figure1}(a). The sample surface is illuminated by a X-ray spot of $100~\mu\mathrm{m}$ diameter. The scattered intensity of the structural $(001)$ Bragg peak and its magnetic satellites is captured using a photo diode. Initially the diode angle $(2\theta)$ is adjusted such that the intensity of the Bragg peak is maximized. Subsequently, $(2\theta)$ remains fixed and the reciprocal space around the Bragg peak is mapped by varying the sample angle $\omega$ and the vertical diode position $z_{d}$. The diode is located behind a pinhole with a diameter of 300~$\mu$m that convolutes the measured signal, leading to an elongation of the peaks in linearly scanned $z_{d}$ direction. Despite the rather small sample--detector distance of 20~mm, the Bragg peak and its magnetic satellites may be separated clearly. Using the superconducting vector magnet at VEKMAG, the magnetic field is applied either parallel to the $[110]$ axis, the $[001]$ axis, or the incoming X-ray beam.

The magnetic phase diagram of Cu$_{2}$OSeO$_{3}$ is schematically depicted in Fig.~\ref{figure1}(b)~\cite{2012_Seki_Science, 2016_Bauer_Book}. Starting in the paramagnetic~(pm) state at high temperatures and low fields, long-wavelength helimagnetic order (wavelength $\lambda = 620$~\AA) is observed below the transition temperature $T_{c} = 58$~K. In the helical state, macroscopic domains of helices propagate along one of the easy $\langle001\rangle$ axes. In the corresponding reciprocal space map, shown in Fig.~\ref{figure1}(c), the structural $(001)$ Bragg peak is surrounded by four magnetic satellite peaks $(\delta01)$, $(\bar{\delta}01)$, $(0\delta1)$, and $(0\bar{\delta}1)$. Although after zero-field cooling the equivalent domains are expected to be populated equally across a bulk sample, the present experiment only maps few domains due to the small X-ray spot size, rendering asymmetric intensity distributions possible.

Under applied magnetic field, the propagation directions of the helices re-orient into the field direction and finite net magnetization emerges as the magnetic moments increasingly tilt towards the field direction. In REXS, this conical state is associated with magnetic satellites along the field direction that are observed when the field is applied perpendicular to the $[001]$ direction, as shown in Fig.~\ref{figure1}(d). Further increasing the magnetic field to the critical field $H_{c2}$ results in a field-polarized state in which the moments are aligned along the field and no long-wavelength modulation is observed (not shown). 

At intermediate magnetic field just below $T_{c}$, a pocket of skyrmion lattice state is observed. The trigonal order of the spin whirls in the plane perpendicular to the field translates to the characteristic sixfold pattern of magnetic satellites in both small-angle neutron scattering~\cite{2012_Adams_PhysRevLett, 2012_Seki_PhysRevB, 2014_White_PhysRevLett, 2017_Bannenberg_PhysRevB, 2017_Makino_PhysRevB} and REXS~\cite{2014_Langner_PhysRevLett, 2016_Zhang_PhysRevB, 2016_Zhang_NanoLett, 2016_Zhang_ApplPhysLett, 2017_Zhang_NatCommun, 2018_Zhang_ProcNatlAcadSciUSA}, as shown in Fig.~\ref{figure1}(e). Under field cooling, the skyrmion lattice may be frozen-in to lower temperatures as a metastable state~\cite{2010_Munzer_PhysRevB, 2016_Okamura_NatCommun, 2017_Seki_PhysRevB, 2018_Bauer_Book}. As depicted in Fig.~\ref{figure1}(f), the intensity of the magnetic satellites increases by an order of magnitude due to the increase of the magnetic moment with decreasing temperature.

\begin{figure}
	\includegraphics[width=0.8\linewidth]{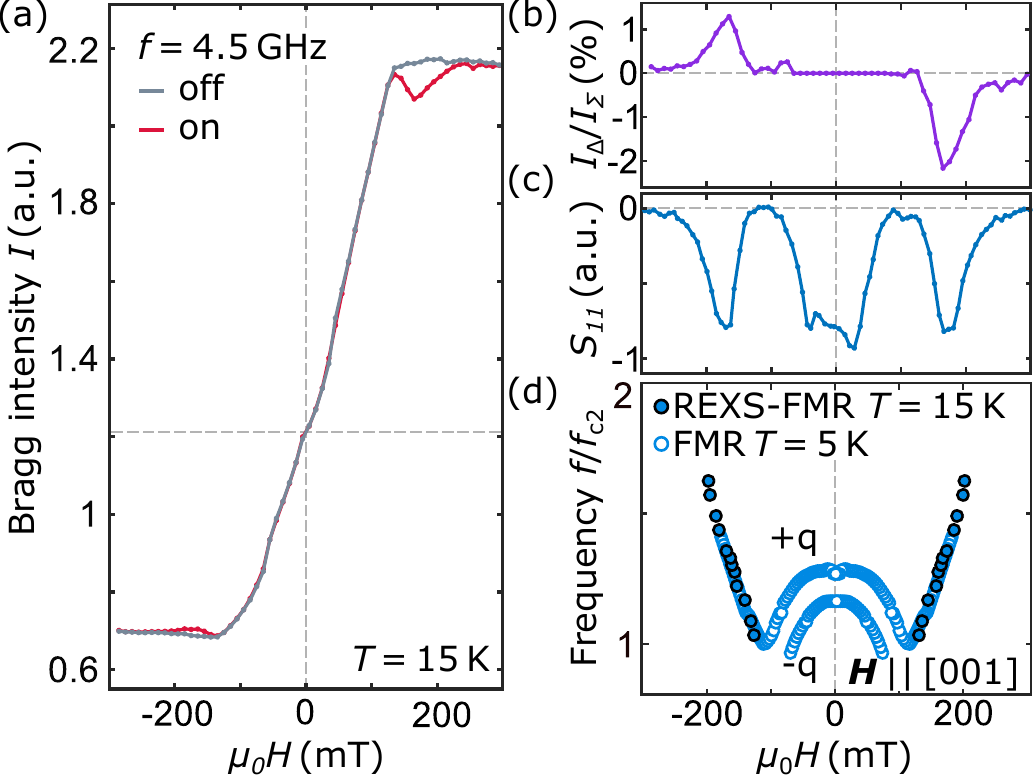}	
	\caption{\label{figure2}REXS-FMR on the structural $(001)$ peak. (a)~Intensity of the Bragg peak as a function of magnetic field applied parallel to the incoming X-ray beam, tracking the magnetization of the sample. Data are recorded at each field step with microwave excitation on (red curves) and off (gray curves). (b)~Relative change of the intensity under excitation. (c)~Reflected microwave power, $S_{11}$, as a function of field. The signature around zero field is attributed to the helimagnetic states. (d)~Resonant modes for magnetic field along the $[001]$ axis at low temperature. Frequencies are normalized to their value at $H_{c2}$. Resonance fields inferred from REXS-FMR (solid symbols) are compared to resonance frequencies inferred from microwave spectroscopy (open symbols).}
\end{figure}

Typical REXS-FMR data are shown in Fig.~\ref{figure2} for the field-polarized state at $T = 15$~K, when a field larger than $\mu_{0}H_{c2} \approx 125$~mT is applied parallel to the beam direction \footnote{Note that critical field values slightly vary with the applied field direction in a bar-shaped sample due to demagnetization effects.}. Due to the lack of a long-wavelength modulation there are no magnetic satellites around the structural $(001)$ Bragg peak. This peak, however, comprises a magnetic contribution that depends on the magnitude and orientation of the magnetization $\vec{M}$ with respect to the incident and scattered X-ray wave vector. In turn, the field dependence of the magnetization may be inferred by tracking the intensity of the Bragg peak, shown in Fig.~\ref{figure2}(a). Clear kinks at $\pm H_{c2}$ and saturated behavior at larger fields are observed, when the microwave excitation is switched off (gray curve). Subtle changes in the curve's slope indicate phase transitions which are discussed in more detail in Refs.~\cite{2012_Adams_PhysRevLett, VersteegVergaraSchaeferEtAl2016, 2018_Halder_PhysRevB}. As the magnetic contribution includes terms linear and quadratic in $\vec{M}$, the intensity curve is not symmetric with respect to zero field. We refer to the Supplementary Material~\cite{} and Refs.~\cite{2008_vanderLaan_CRPhys, 2017_Zhang_PhysRevB} for information on the determination of the absolute magnetization values.

Resonant excitation is studied by repeatedly switching on and off the microwave excitation while stepping the magnetic field, starting in high positive fields. At each field point, the REXS intensity under excitation is integrated for 3~s before the excitation is switched off and the intensity is integrated again for 3~s. For an excitation frequency of 4.5~GHz (red curve), a minimum of the magnetic intensity contribution emerges in the field-polarized state above $H_{c2}$. As shown in Fig.~\ref{figure2}(b), the normalized signal difference, $(I_{\mathrm{on}}-I_{\mathrm{off}}) / (I_{\mathrm{on}}+I_{\mathrm{off}})$, at this minimum is of the order of 2\%, which translates to a reduction of the magnetization by about 6.5\%. Note that the way how the excitation is applied means that the magnetization reproducibly switches from its reduced to its regular value at each field step. When assuming that precessional motion of the moments causes the reduction, a precession angle of $21^{\circ}$ is required. This value is large but plausible considering the rather low effective damping of $\alpha\approx 10^{-4}$ observed in the insulator Cu$_{2}$OSeO$_{3}$~\cite{2017_Stasinopoulos_ApplPhysLett, 2015_Bauer_NatCommun}. In contrast, a change of the moment of 6.5\% by heating effects requires an increase/decrease of the sample temperature by ${\sim}13$~K after switching the excitation on/off in less than the time frame of 1~s resolvable in the present REXS experiment. Such drastic heating effects, however, would interfere when studying the skyrmion lattice state with its rather narrow temperature width of ${\sim}2$~K close to $T_{c}$, see below, and can be excluded. 

Simultaneously to the REXS measurements, the reflected microwave power $S_{11}$ of the coplanar waveguide was recorded using a Schottky diode detector. As shown in Fig.~\ref{figure2}(c), minima in the field-polarized state above $\pm H_{c2}$ are observed at the same field values as in REXS. An additional broad signature around zero field is attributed to resonant excitations in the helimagnetic state, notably of the $\pm$q modes. The absence of these resonances in the REXS data highlights the potential of REXS-FMR to selectively study individual magnetic phases and determine the origin of specific excitations. Figure~\ref{figure2}(d) shows that with increasing excitation frequency the resonance field in REXS data increases linearly, consistent with Kittel behavior in the field-polarized state. The resonance field values in the field-polarized state are also in excellent agreement with resonance frequencies inferred from conventional microwave spectroscopy using a vector network analyzer (spectra not shown).

\begin{figure}
	\includegraphics[width=0.52\linewidth]{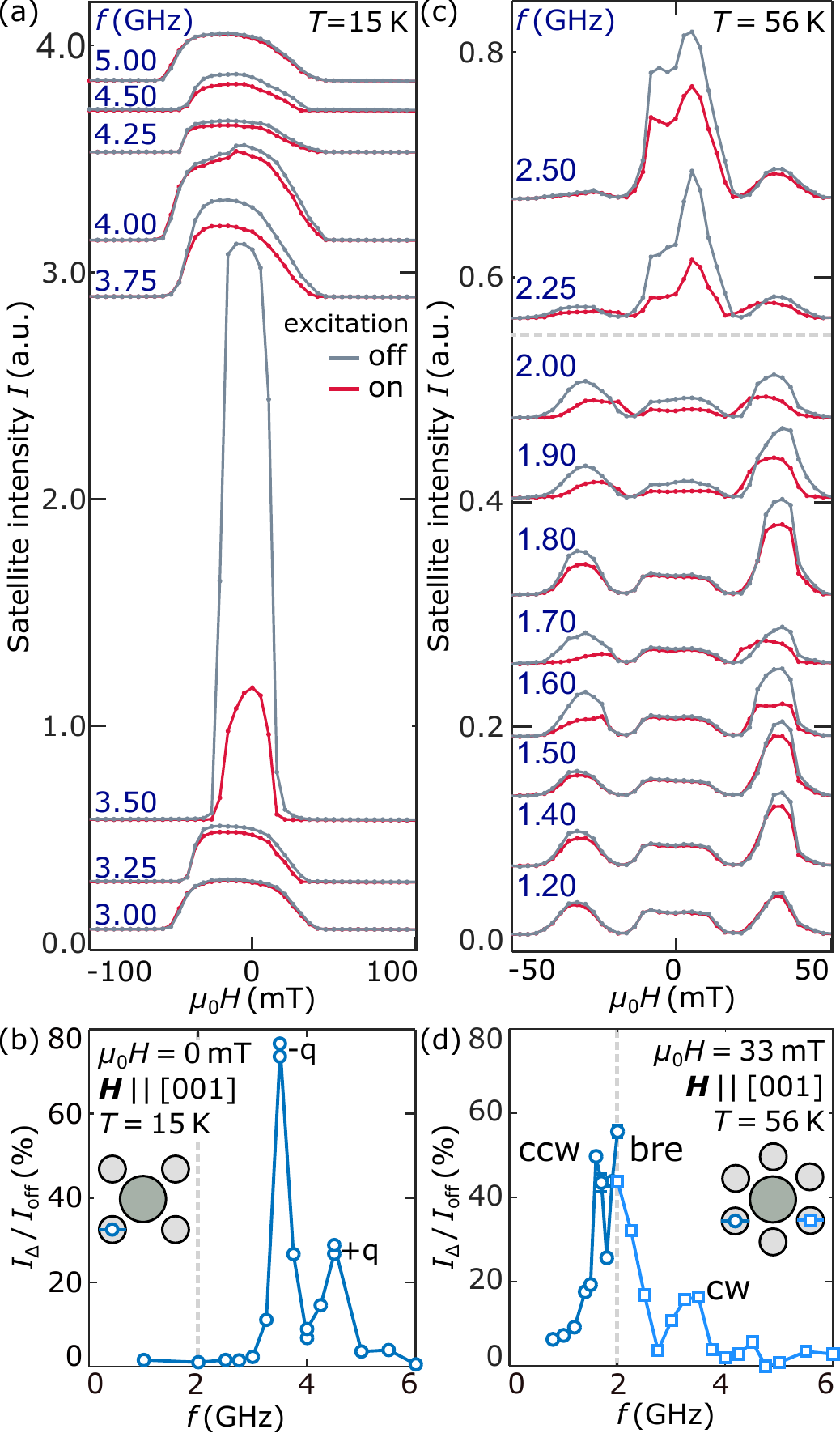}	
	\caption{\label{figure3} REXS-FMR on the magnetic satellites. (a)~Intensity of a helical satellite as a function of magnetic field for different excitation frequencies. Data are recorded at each field step with microwave excitation on (red curves) and off (gray curves). The large value at 3.5~GHz is attributed to changes of the spin texture, see text for details. (b)~Relative reduction of the helical satellite intensity in zero field as a function of frequency. The sketch illustrates the position of the detector diode. Statistical error bars are typically much smaller than the symbol size. (c)~Intensity at the reciprocal space position of a skyrmion lattice satellite. Around zero field, the tail of the broad helical satellite is observed at the skyrmion satellite position. (d)~Relative reduction of the skyrmion satellite intensity at a field of $35$~mT as a function of frequency.}
\end{figure}

As one of its decisive advantages, REXS-FMR may be carried out not only on structural Bragg peaks but also on magnetic satellites, allowing to unambiguously make the connection between the underlying magnetic phase and the resonant mode. In Fig.~\ref{figure3}(a), the intensity at a helical satellite position is shown as a function of field for different excitation frequencies at $T = 30$~K. Similar as before, data are recorded with microwave excitation switched on and off at each magnetic field step. The integration time was increased to 10~s. Finite intensity arises only around zero field, where the helical state is observed in the magnetic phase diagram after zero-field cooling, cf.\ Fig.~\ref{figure1}(b). Note, however, that the field is applied along $[001]$, i.e., an easy axis for the helices in Cu$_{2}$OSeO$_{3}$, and the measurement starts in the field-polarized state. Therefore, when decreasing the field to zero through the conical state, the helices may be expected to remain in the helical domain oriented along the field direction even in zero field~\cite{2017_Bauer_PhysRevB}, resulting in the absence of magnetic satellites in the present scattering geometry at temperatures well below $T_{c}$.

This putative contradiction connects to the recent discovery that Cu$_{2}$OSeO$_{3}$ not only hosts a skyrmion lattice at high temperatures, common to all cubic chiral magnets, but also an independent second skyrmion phase at low temperatures~\cite{2018_Chacon_NatPhys, 2018_Halder_PhysRevB}. In contrast to the high-temperature phase, the low-temperature phase stabilized by magneto-crystalline anisotropies and exists only for field values around $H_{c2}$ applied along $[001]$. Due to the topological protection inherent to skyrmions and the rather low temperature, the energy barrier of the low-temperature skyrmion phase is comparatively high.
As a result, the skyrmion state exhibits a rather glassy texture without well-defined long-range order. When this state decays at lower fields by means of coalescence of neighboring skyrmions, a texture resembling poorly ordered helices with propagation perpendicular to the field forms~\cite{2013_Milde_Science, 2017_Wild_SciAdv}. The weak magnetic satellites associated with such a helical state are detected in our REXS experiment. Perhaps most strikingly, the glassy texture is highly susceptible to changes induced by resonant microwave excitation, as explained in the following. 

At low and high frequencies, i.e., when the excitation is off-resonance, data with microwave switched on and off agree with each other, corroborating that heating effects are negligible. In resonance, the excitation distinctly reduces the satellite intensity. Consistent with the behavior in the field-polarized state, the intensity reproducibly switches between its low and high value at each field step and the reduction is attributed to a precessional motion of the magnetic moments. Note, however, that data without excitation (gray curves) are expected to track each other as long as the same magnetic texture is probed, which is clearly not the case.

The discrepancy becomes especially pronounced for an excitation frequency of 3.5~GHz, for which the satellite intensity increases by an order of magnitude while the field range shrinks by a factor of two. This finding suggests that the microwave excitation interacts with the glassy magnetic texture described above, improving its long-range order. In resonance this pumping effect is particularly effective and the corresponding magnetic satellite in reciprocal space not only increases in intensity but also decisively sharpens.

When also taking into account a small misalignment between sample plane and field direction, as a result the scattering condition is only fulfilled in a reduced field interval around zero field. In Fig.~\ref{figure3}(b), the relative reduction of the helical satellite intensity in zero field is shown as a function of the excitation frequency. Two sharp maxima are observed that are unambiguously attributed to the -q and +q collective modes of the helical state. Both frequency values and heights of the maxima are in excellent agreement with the literature~\cite{2012_Onose_PhysRevLett, 2015_Schwarze_NatMater}. 

Finally, REXS-FMR on the high-temperature skyrmion lattice is presented in Fig.~\ref{figure3}(c) showing the intensity at a skyrmion satellite position as a function of field at $T = 56$~K for different excitation frequencies. Starting the description in off-resonance at low excitation frequency (bottom), measurements with and without microwave excitation track each other. The intensity maxima in finite fields around $\pm30$~mT are attributed to the skyrmion lattice state. In addition, finite intensity emerges around zero field as the tail of the broad helical satellite close-by in reciprocal space reaches the position of the skyrmion satellite. At higher frequencies, a distinct reduction of the skyrmion satellite intensity is observed under microwave excitation, again attributed to a precessional motion of the magnetic moments. At frequencies above 1.9~GHz, resonance effects associated with the helical state are observed around zero field.  

In order to further analyze the skyrmion resonances, the relative reduction of the satellite intensity at 35~mT, i.e., in the skyrmion lattice state, is depicted in Fig.~\ref{figure3}(d) as a function of the excitation frequency. Three distinct maxima are observed and associated with the counter-clockwise~(ccw) gyration, breathing~(bre), and clockwise~(cw) gyration mode, in excellent agreement with literature~\cite{2012_Onose_PhysRevLett, 2015_Schwarze_NatMater}. The relative heights of the maxima are also perfectly consistent with calculated spectral weight distributions, as the coplanar waveguide combines in-plane excitation, driving the two gyration modes, and out-of-plane excitation, driving the breathing mode~\cite{2017_Stasinopoulos_SciRep}. Due to a photon penetration depth of about 30~nm, the REXS-FMR measurements probe the recently discovered surface states of the skyrmion lattice in Cu$_{2}$OSeO$_{3}$~\cite{PhysRevLett.120.227202, LegrandChauleauMaccarielloEtAl2018}. Although roughly 25~\% of the probed volume are expected to contain such surfaces states, no contributions distinguishable from the bulk resonance modes are observed, indicating that the resonance frequencies of bulk and surface states are very similar or equal. Note that above and below 2~GHz different circulators are used, leading to a small offset in the applied microwave power. Furthermore, the diode position was changed, as indicated in the sketch in Fig.~\ref{figure3}(d), resulting in quantitative discrepancies in the measured intensity profiles at that frequency.

In summary, we established a novel X-ray scattering technique, referred to as REXS-FMR, that combines microwave excitation by means of a coplanar waveguide with detection in reciprocal space by means of magnetic REXS. In the cubic chiral magnet Cu$_{2}$OSeO$_{3}$, we identified the resonant modes in the field-polarized, helimagnetic, and skyrmion lattice state by tracking the intensity of the structural $(001)$ Bragg peak and its magnetic satellites under microwave excitation. The surface sensitive measurement indicates equal magnetic resonance frequencies for skyrmionic surface and bulk states. REXS-FMR also allows for stroboscopic measurements in order to determine the character of eigenmodes microscopically, e.g., to distinguish between gyration and breathing modes. Also note that the selectivity to certain magnetic phases may prove particularly useful in complex magnetic environments for which the clear identification of eigenmodes in conventional microwave spectroscopy is challenging, such as for multi-domain skyrmion states in lacunar spinels~\cite{2015_Kezsmarki_NatMater, 2016_Ehlers_PhysRevB, 2019_Okamura_PhysRevLett}, glassy skyrmionic textures in Co-Mn-Zn compounds~\cite{2015_Tokunaga_NatCommun, 2017_Takagi_PhysRevB, 2018_Karube_SciAdv}, or the low-temperature skyrmion state in Cu$_{2}$OSeO$_{3}$ with its concomitant tilted conical state~\cite{2018_Chacon_NatPhys, 2018_Halder_PhysRevB, 2018_Qian_SciAdv}.

We wish to thank W.\ Simeth for fruitful discussions and assistance with the experiments.
C.B.\, G.W.\ and F.R.\ acknowledge funding by the BMBF via the VEKMAG project.
S.P.\, G.W.\ and C.B.\ acknowledge funding by the German Research Foundation via SPP2137.
This project has received funding from the EMPIR programme co-financed by the Participating States and from the European Union's Horizon 2020 research and innovation programme.
A.B.\ and C.P.\ acknowledge financial support through DFG TRR80 (project F7), DFG SPP2137, as well as ERC Advanced Grants 291079 (TOPFIT) and 788031 (ExQuiSid). This work has been funded by the Deutsche Forschungsgemeinschaft (DFG, German Research Foundation) under Germany’s Excellence Strategy – EXC-2111 – 390814868.

%

\clearpage

\textit{The citations in this appended supplement refer to the ones in the main text given above.}

\begin{center}\large{\bf{Ferromagnetic resonance with magnetic phase selectivity by means of resonant elastic x-ray scattering on a chiral magnet - supplementary material}}
	\end{center}

When the Bragg-peak intensity is recorded as a function of externally applied magnetic field a magnetization trace is observed. The observed intensity change is however not directly proportional to the change of magnetization due to higher order terms. To extract the actual magnetization trace the theory of van der Laan et al. is applied in the following \cite{2008_vanderLaan_CRPhys}.

The scattering cross section $I(\mathbf{q})$ is given by (\cite{2008_vanderLaan_CRPhys} Equation (15))

\begin{equation}
I\left(\mathit{\mathbf{q}}\right)=I_c \left(\mathit{\mathbf{q}}\right)+I_m \left(\mathit{\mathbf{q}}\right)+I_i \left(\mathit{\mathbf{q}}\right) \label{I}
\end{equation}

with the scattering vector $\mathbf{q} = \mathbf{k}_f-\mathbf{k}_i$, the incident and final photon wave vector $\mathbf{k}_i$ and $\mathbf{k}_f$ (and their unit vectors $\mathbf{\hat{k}}_i$ and $\mathbf{\hat{k}}_f$). The contribution of charge scattering $I_c$ is given by (\cite{2008_vanderLaan_CRPhys} Equation (16))

\begin{equation}
I_c(\mathbf{q}) = \frac{1}{2} \left|\tilde{f}_0 \tilde{\rho}\right|^2 \left[P_\sigma + P_\pi \left|\mathbf{\hat{k}}_f \cdot  \mathbf{\hat{k}}_i\right|^2  \right] \label{Ic}
\end{equation}

with the Fourier transformation of the charge density $\tilde{\rho}$ and the monopole contribution of the energy dependent resonance amplitude $\tilde{f}_0$. Further is $P_\sigma = P_0 + P_1$ and $P_\pi = P_0-P_1$ with the Poincar\'e vector $\mathbf{P}$. For circularly polarized light $\mathbf{P}^T = (1,0,0,\pm 1)$ which is assumed for the following considerations. Equation \ref{Ic} is independent of the magnetization $\mathbf{M}$ (and its Fourier transformation $\mathbf{\tilde{M}}$) of the sample.

The contribution of pure magnetic scattering $I_m$ for $\tilde{\mathbf{M}}$ inside the scattering plane is given by (\cite{2008_vanderLaan_CRPhys} Equation (19))

\begin{equation}
I_m(\tilde{\mathbf{M}}_{q\parallel}) = \frac{1}{2} \left| \tilde{f}_1 \right|^2 \left[ P_\sigma \left| \mathbf{\hat{k}}_f \cdot \mathbf{\tilde{M}} \right|^2 + P_\pi \left| \mathbf{\hat{k}}_i \cdot \mathbf{\tilde{M}} \right|^2 \right]
\end{equation}

with the Fourier transformation of the magnetic dipolar contribution to the energy dependent resonance amplitude $\tilde{f_1}$. Finally, the interference term between charge and magnetic scattering $I_i$ for circular polarization with $\tilde{\mathbf{M}}$ inside the scattering plane is given by (\cite{2008_vanderLaan_CRPhys} Equation (21))

\begin{align}\label{Ii}
I_i(\tilde{\mathbf{M}}_{q\parallel}) = 
P_2 \text{Im} \left[ \tilde{f}_0^*  \tilde{\rho}^* \tilde{f}_1  \tilde{\rho}^* \left( \mathbf{\hat{k}}_i \cdot \mathbf{\tilde{M}} + (\mathbf{\hat{k}}_f \cdot \mathbf{\tilde{M}} )(\mathbf{\hat{k}}_f \cdot \mathbf{\hat{k}}_i ) \right) \right] \\ \nonumber +
P_3 \; \text{Re}{\left[ \tilde{f}_0^*  \tilde{\rho}^* \tilde{f}_1^* \left( \mathbf{\hat{k}}_i \cdot \mathbf{\tilde{M}} + (\mathbf{\hat{k}}_f \cdot \mathbf{\tilde{M}} )(\mathbf{\hat{k}}_f \cdot \mathbf{\hat{k}}_i ) \right) \right]}  
\end{align}

For the case shown in Fig. 2(a) of the main text, the magnetization is collinear to the incident beam. The photon wave vectors $\mathbf{k}_i$ and $\mathbf{k}_f$ are fixed to the (001)-Bragg condition ($2\theta = 95.8^\circ$). As a function of $\mathbf{M}(\mu_0 H_\text{ext})$ the charge scattering $I_c = A_c$ is constant. $\mathbf{P}$ is assumed for circularly polarized light, meaning $P_\sigma = P_\pi = 1$, $P_2=0$ and $P_3=\pm 1$. 

\begin{figure}[p]
	\includegraphics[width = \linewidth]{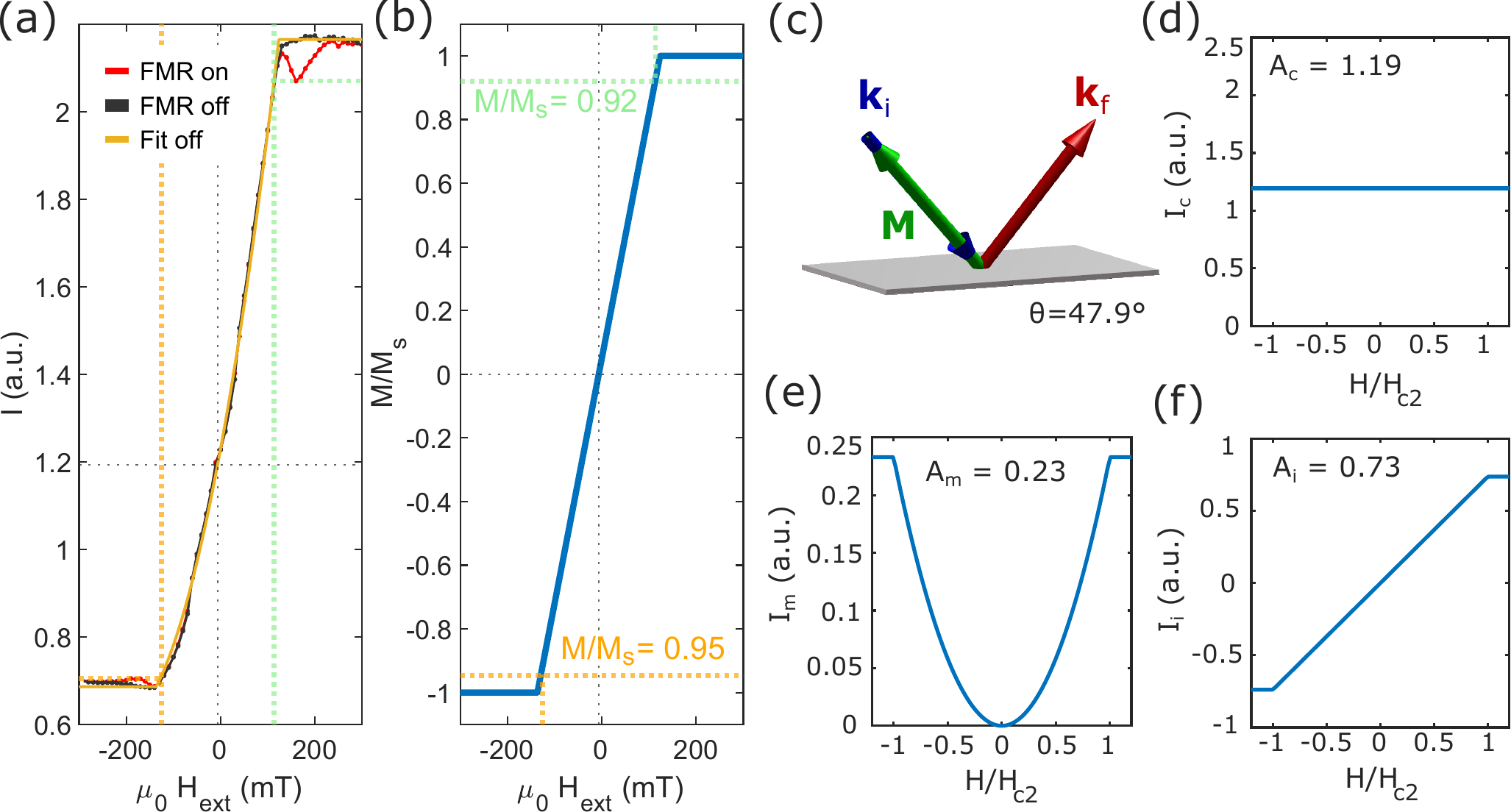}	
	\caption{\label{fig_fmr_fp_supp} \textbf{(a)} Magnetization trace obtained by tracking the Bragg peak intensity with applied magnetic field (black). The red curve shows the same trace with an applied microwave excitation of 4.5~GHz (Compare Fig. 2(a) of the main text). Also included is a fit to the theoretical formula (orange). \textbf{(b)} Magnetization as a function of externally applied magnetic field obtained from fit. \textbf{(c)} Scattering geometry present in (a). \textbf{(d-f)} Contribution of charge, pure magnetic and interference scattering to the Bragg peak intensity. } \label{fig_S1}
\end{figure}

Equation \ref{I} can then be written as

\begin{align} \label{fitcurve}
I(\tilde{\mathbf{M}}_{q\parallel}) = A_c + A_m \left(  \left| \mathbf{\hat{k}}_f \cdot \mathbf{\tilde{M}} \right|^2 + \left| \mathbf{\hat{k}}_i \cdot \mathbf{\tilde{M}} \right|^2 \right) \pm A_i \left( \mathbf{\hat{k}}_i \cdot \mathbf{\tilde{M}} + (\mathbf{\hat{k}}_f \cdot \mathbf{\tilde{M}} )(\mathbf{\hat{k}}_f \cdot \mathbf{\hat{k}}_i ) \right) 
\end{align}

with the constants $A_m = \frac{1}{2} \left| \tilde{f}_1 \right|^2 M_s$ and 
$A_i = \text{Re}\left[ \tilde{f}_0^*  \tilde{\rho}^* \tilde{f}_1^*\right] M_s$. The net magnetization $\mathbf{M}(\mu_0 H_\text{ext})$ along $\mathbf{\hat{k}}_i$ is assumed to have the following form

\begin{align} \label{M}
\mathbf{M}(\mu_0 H_\text{ext}) = \begin{cases}  
\mathbf{\hat{k}}_i \frac{H_\text{ext}}{H_\text{c2}} M_s, & \text{for $\left|H_\text{ext} \right| < H_\text{c2}$} \\
\mathbf{\hat{k}}_i M_s, & \text{otherwise}
\end{cases}
\end{align}

with the second critical field $H_\text{c2}$ and the saturation magnetization $M_s$. For $\left|H_\text{ext} \right| < H_\text{c2}$ this is obviously not true in case of the non-collinear ferromagnet that is investigated, the net magnetization along the field however is well-described by this behaviour as can be seen e.g. from \cite{2012_Adams_PhysRevLett}. The experimental data in Fig. 2(a) of the main text can now be fitted to equation \ref{fitcurve} using the parameters $A_c$, $A_m$, $A_i$ and $H_{c2}$. Fig.~\ref{fig_S1}(a) shows again the data from Fig. 2(a) of the main text and the resulting fit for the excitation off data. The fit gives a ratio of pure magnetic scattering to the interference term of $\frac{A_i}{A_m}=3.2$. From the fit parameter $H_\text{c2}$ the magnetization trace can be plotted using equation \ref{M}. Getting $H_\text{ext}$ from the plot where the intensity of the rf off data (in the non-saturated part) is equal to the reduced intensity of the rf on data, one can use Fig.~\ref{fig_S1}(b) to estimate the reduction $\Delta M$ of $M$ in resonance. A mean value of $\Delta M / M_s \approx 0.065$ is obtained. The scattering geometry is shown in Fig.~\ref{fig_S1}(c). The contributions of charge scattering, pure magnetic and interference scattering obtained from the fit is shown in Fig.~\ref{fig_S1}(d-f).

\end{document}